\documentclass{aa}
\usepackage{natbib,psfig,graphicx}
\usepackage{txfonts}
\def\a26{Abell~2667}

\begin{document}


\title{Diffuse light and building history of the galaxy cluster Abell~2667}

\author{
G. Covone\inst{1,2} \and C. Adami\inst{2} \and F. Durret\inst{3,4} 
\and J.-P. Kneib\inst{2} \and G.B. Lima Neto\inst{5}
\and E. Slezak\inst{6} }

\offprints{C. Adami \email{christophe.adami@oamp.fr}}

\institute{
INAF -- Osservatorio Astronomico di Capodimonte, Naples, Italy
\and
Laboratoire d'Astrophysique de Marseille, 
Traverse du Siphon, 13012 Marseille, France
\and
Institut d'Astrophysique de Paris, CNRS, Universit\'e Pierre et Marie Curie,
98bis Bd Arago, F-75014 Paris, France 
\and
Observatoire de Paris, LERMA, 61 Av. de l'Observatoire, 75014 Paris, France
\and
Instituto de Astronomia, Geof\'{\i}sica e C. Atmosf./USP, R. do Mat\~ao 1226, 
05508-090 S\~ao Paulo/SP, Brazil
\and
Observatoire de la C\^ote d'Azur, B.P. 4229, 06304 Nice Cedex 4, France
}

\date{Accepted . Received ; Draft printed: \today}

\authorrunning{G. Covone et al.}
\titlerunning{Diffuse light in Abell~2667}

\abstract
{}
{We have searched for diffuse intracluster light in the 
galaxy cluster Abell~2667 ($z=0.233$) from HST
images in three broad band filters.}
{We have applied to these images an iterative multi-scale wavelet
analysis and reconstruction technique, which allows to subtract stars
and galaxies from the original images.}
{We detect a zone of diffuse emission south west of the cluster center
(DS1), and a second faint object (ComDif), within DS1. Another diffuse
source (DS2) may be detected, at lower confidence level, north east of
the center. These sources of diffuse light contribute to 10-15$\%$ of
the total visible light in the cluster. Whether they are independent
entities or are part of the very elliptical external envelope of the
central galaxy remains unclear.  Deep VLT VIMOS integral field
spectroscopy reveals a faint continuum at the positions of DS1 and
ComDif but do not allow to compute a redshift and conclude if these
sources are part of the central galaxy or not.  A hierarchical
substructure detection method reveals the presence of several galaxy
pairs and groups defining a similar direction as the one drawn by the
DS1 - central galaxy - DS2 axis.  The analysis of archive XMM-Newton
and Chandra observations shows X-ray emission elongated in the same
direction. The X-ray temperature map shows the presence of a cool
core, a broad cool zone stretching from north to south and hotter
regions towards the north east, south west and north west. This
possibly suggests shock fronts along these directions produced by
infalling material, even if uncertainties remain quite large on the
  temperature determination far from the center.}
{These various data are consistent with a picture in which
diffuse sources are concentrations of tidal debris and harassed matter
expelled from infalling galaxies by tidal stripping and undergoing an
accretion process onto the central cluster galaxy; as such, they are
expected to be found along the main infall directions. Note
however that the limited signal to noise of the various data and the
apparent lack of large numbers of well defined independent tidal
tails, besides the one named ComDif, preclude definitive conclusions
on this scenario.} 


\maketitle

\section{Introduction}
\label{intro}

The dense environment of galaxy clusters has a strong influence on
galaxy evolution.  Galaxies which enter galaxy clusters are affected
by a large variety of physical processes during their interactions
with the individual cluster galaxies, the intra-cluster medium (ICM)
and the whole cluster gravitational field.  However, the details and
the relative role of these interactions at different epochs and in
different cluster regions are not yet fully understood (see, e.g.,
Treu et al. 2003).  Some of these processes are held responsible for
removing gas from the disk of spiral galaxies (ram pressure stripping,
see e.g. Gunn \& Gott 1972), the halo gas reservoir (tidal truncation
of galactic halos; see, e.g., Natarajan, Kneib \& Smail 2002),
generally leading to a decreasing star formation rate ({\em
starvation}, Larson et al. 1980).

Multiple high-speed encounters ({\em harassment}, Moore et al. 1998)
are, on the other side, responsible for removing stars from the disk
of late-type galaxies and creating tidal debris.  Numerical
simulations (e.g., Calc\'aneo-Rold\'an et al. 2000, Napolitano et
al. 2003, Sommer-Larsen et al. 2005) have shown that this might be the
main mechanism leading to the formation of a population of
intracluster stars bound not to individual galaxies but to the cluster
as a whole.  This population is now detected as a faint, diffuse
optical source, the intracluster light (ICL).  Moreover, such stellar
population is also expected to be a relevant remnant of the merging
events which lead to the formation of the central bright galaxy (e.g.,
Lin \& Mohr 2004).  Therefore, the study of the ICL is expected to
give important information on the physical processes transforming the
galaxies accreted in dense enviroments and on the formation of the
cluster brightest galaxies and their extended envelope.

The stellar population responsible for the ICL
may be probed both by observations of the 
individual stars
(see, for instance, the recent review in Arnaboldi 2004)
and of the cumulative diffuse light.
Deep imaging surveys are accumulating a growing and detailed
evidence
for diffuse structures with various shapes in several clusters 
(e.g., Gregg \& West 1998, Feldmeier et al. 2002, 2003, Mihos 2003, 
Adami et al. 2005, Da Rocha \& Mendes de Oliveira 2005, Krick et al. 2006),
employing a variety of techniques
to face the challenging task to detect a diffuse source 
with a surface brightness that
can be as low as $\sim 1$ \% of the night sky level.

Another known difficulty is given by the connection between the ICL
and the brightest cluster galaxy envelope (see for instance Uson et
al. 1991): this has led to various definitions of ICL in literature.
Therefore, some care is needed when comparing different estimates of
the amount of light found in the ICL population and those obtained
from numerical simulations.  In the following we assume that the
brightest cluster galaxy envelope has the same origin and nature as
the population of stars bound to the cluster as a whole, and make no
strict distinction between them.

Lin $\&$ Mohr (2004) have studied the near-infrared properties of a
sample of 93 nearby, X-ray selected clusters and groups, showing that
ICL is a common feature, accounting for a large part of the present
total amount of stars (up to 50$\%$ for $\geq$10$^{15}\ M_\odot$
clusters).  The ICL contribution appears to increase with the total
cluster mass.  Note that in this work no distinction has been
made between the ICL and the brightest cluster galaxy envelope.

In a sample of 24 clusters at $z < 0.13$, Gonzalez et al. (2005) have
used a two-component $r^{1/4}$ profile to fit the central elliptical
galaxy and the population of intracluster stars.  The outer component
is found to be distributed on scales $\sim 10-40$ times larger than
the inner one, and $\sim$ 10 times more luminous.
Moving at higher redshifts, Zibetti et al. (2005) have shown evidence
for the presence of ICL in rich galaxy clusters in the redshift
range 0.2-0.3.  They have stacked a sample of 683 galaxy clusters from
the Sloan Digital Sky Survey and found that the ICL contribution to
the total cluster light in the stacked cluster is 10.9$\pm$5.0\%, that
it shares similar colors with the cluster population (within the
photometric errors) but has a higher central concentration.

Numerical simulations have also greatly improved the description of
the behavior of the diffuse light component.  For instance, Napolitano
et al. (2003), by using a high-resolution collisionless simulation,
have found that in a Virgo-like cluster the velocity distribution of
the ICL stars is strongly non-Gaussian also at $z=0$, constituted of a
large spectrum of structures (filaments, concentrations of particles,
etc...) and showing a clearly unrelaxed state, with a typical
clustering radius of about 50 kpc.
Murante et al. (2004) confirm with numerical simulations that the ICL
contribution to the stellar mass seems to increase with the cluster
mass and represents more than 10$\%$ of the total amount of stars.
They also found the ICL stars to be older than galactic stars.

Sommer-Larsen et al. (2005) find an intracluster star contribution
to the total cluster light in the B band of 20 to 40$\%$ at $z=0$, 
with a B-R color of 1.4-1.5
and a metallicity varying from solar to half solar.

\begin{figure}
\centering
\caption{WFPC2 image of \a26\ in the F606W filter. 
The thick box indicates the 
VIMOS-IFU field of view ($54 \times 54$ arcsec$^2$). 
The overlayed grid shows the North orientation towards the upper left.}
\label{im_raw}
\end{figure}

The number of clusters with individual well studied diffuse light
sources is, however, still quite small and, in a large fraction,
limited to the local Universe.  In order to test predictions from the
numerical simulations at different cosmological epochs, it is
therefore mandatory to enlarge this limited sample.

In this framework, we studied the moderately distant galaxy 
cluster, Abell~2667 ($z = 0.233$).
Abell~2667 is a known gravitational lensing
cluster, with several multiple image systems, including a bright giant
gravitational arc with the source located at $z=1.0328$ (Covone et
al. 2006a). 
This is a massive X-ray luminous galaxy cluster: within the range
[0.1 - 2.4] keV, its X-ray luminosity is estimated to be $ L_X =
(14.90 \pm 0.56) \times10^{44} \, h_{70}^{-2}$ erg s$^{-1}$ (Rizza et
al.  1998). 
Its X-ray morphology (Rizza et al. 1998) and galaxy
dynamical analysis (Covone et al. 2006a) support the idea that its inner
core is relaxed, while wide field imaging shows that the cluster is
still accreting galaxies and galaxy groups (H. Ebeling, private
communication).  
Recently, Covone et al. (2006a) have performed a spectroscopic survey of the
cluster core by using the VIMOS Integral Field Unit (IFU),
and compared the mass models obtained 
through gravitational lensing, cluster galaxy dynamics and X-ray gas
temperature.  
The spectroscopically confirmed galaxy population in the cluster core
is made of evolved galaxies, with no detectable sign (in the optical
bands) of current or very recent star formation.  The brightest
cluster galaxy shows a bright region with emission lines ([OII],
H$\beta$, [OIII], H$\alpha$), extended along the NE-SW direction,
thought to be associated with a cooling flow (Rizza et al. 1998, see
also Covone 2004).

Our aims are: 
(i) to probe diffuse light physics in an earlier
stage than most of the existing works, as Abell~2667
is located at a redshift of 0.233,
and (ii) to use the detected ICL sources 
to describe and discuss the building history of this cluster,
by means of a new dynamical analysis of the cluster galaxies 
(based on the Covone et al. (2006a) redshift catalog)
and archive X-ray data.

The paper is organized as follows: the optical data and analysis
methods are presented in Sect.~\ref{data}.  Imaging and spectroscopic
results on the ICL are described in Sect.~\ref{images}.  X-ray
observations and results are presented in Sect.~\ref{xrays}, and a
possible building scenario for Abell~2667 is discussed in
Section~\ref{discussion}. Conclusions are drawn in
Section~\ref{concl}.
Throughout this paper we will assume a
cosmological model with $\Omega_{\rm m}=0.3$, $\Omega_{\Lambda}=0.7$
and $H_{0}=70 \, {\rm km} \, {\rm s}^{-1}$\,Mpc$^{-1}$.  At
$z=0.233$ the angular scale is thus 3.722 kpc/arcsec.  Magnitudes are
given in the AB system.

\section{Optical data and analysis}
\label{data}

\subsection{The optical data set}

The imaging and spectroscopic data used in this work have been presented 
in Covone et al. (2006a). We refer to that paper for details on the data
reduction and only present here the most relevant properties.

Detection of diffuse light has been performed on archive 
Hubble Space Telescope ({\it HST}) WFPC2 images. 
These are three broad band filters:  
F450W  (exposure time 12000~s), F606W (4000~s) and F814W (4000~s). 
Using the IRAF package {\tt drizzle}, we have obtained a 
pixel scale of 0.05~arcsec per pixel.
The photometric properties of the given multiband dataset
are summarized in Table~\ref{table:HST}.
An image of the central region of \a26\ is displayed in Fig.~\ref{im_raw},
together with the VIMOS-IFU field-of-view (f.o.v.) 
which covers a region of
about $220 \times 220 \, {\rm kpc}^2$ around the central galaxy.

\begin{table}
\caption{Properties of the WFPC2 data (AB system). }
\begin{center}
\begin{tabular}{lccc}
\hline
\hline
filter & exp time & zero point & sky level \\
       & s        & mag        & mag/arcsec$^{2}$  \\
\hline
\noalign{\smallskip}
F450W & 12000  & 21.94  &  23.44 \\
F606W & ~4000  & 22.99  &  23.09 \\
F814W & ~4000  & 22.08  &  22.39 \\
\hline
\end{tabular}
\end{center}
\label{table:HST}
\end{table}

\begin{table*}
\caption{Properties of the ICL sources detected in \a26.
We list:
J2000 coordinates of the light centroid,
total magnitudes, extension (in the F814w filter), 
average surface brightness (mag/arcsec$^2$) and
detection levels in the three given filters.
Magnitudes are given in the AB system.  }
\begin{center}
\label{table:DS}
\begin{tabular}{lccccccccccc}
\hline
\hline
Source &$\alpha$& $\delta$&  B$_{450}$ & V$_{606}$ & I$_{814}$ & area  
& $\mu_{450}$ & $\mu_{606}$ & $\mu_{814}$  
& F606W & F814W \\
& (degree) & (degree)& & & & (arcsec$^2$) & &  & & & \\
\hline
\noalign{\smallskip}
{\em Diffuse sources} & & & & &  & & & & & & \\
DS1 & 357.910 & -26.088 & 21.9 & 20.9  &  19.9  & 137 & 27.2& 26.2& 25.2& 
 1.9$\sigma$ & 2.7$\sigma$ \\
DS2 & 357.920 & -26.079 &  --  & 21.3  &  20.6  & 143 &   --& 26.7& 26.0&     
 1.5$\sigma$ & 2.1$\sigma$ \\
\noalign{\smallskip}
{\em Compact diffuse source} & & & & & & & & & & & \\
ComDif & 357.9096 & -26.0880 & 27.29 & 25.77  &  24.12  & 2.4 & 28.24 & 26.72 & 25.07 & -- & -- \\

\hline
\end{tabular}
\end{center}
\end{table*}

Integral field spectroscopy has been obtained with the 
VIMOS-IFU mounted on the VLT Melipal, as part 
of an integral field spectroscopic
survey of X-ray bright clusters at $z \simeq 0.2$ 
showing strong gravitational lensing features (Covone et al. 2006b).
Observations were performed using the low-resolution blue grism, 
with a fiber diameter of 0.66~arcsec.
The useful wavelength range is between $\sim 3900$ \AA \, and 6800 \AA, 
with spectral resolution  $R \simeq 250$.
The total exposure time was 10800~s, divided in four exposures 
with a $\sim 1$~arcsec offset among them.

\subsection{Image analysis}
\label{images}

In order to detect large scale faint diffuse light emission sources in
the presence of structures (i.e., stars and galaxies) we first need to
remove these small scale objects. 
In order to obtain an image containing only these
objects, we performed a multi-scale analysis of the images, assuming
that the spatial extension of the diffuse sources is in most cases
larger than the typical galaxy scale. 
Diffuse light is searched for after subtraction of all these
smaller scale sources.  
We only analyzed the area covered by the WFPC2-wf3 chip,
since this encloses the cluster central region and 
the VIMOS-IFU f.o.v.
We summarize our method here and refer to Adami et al. (2005) for a
full description. Recently, Da Rocha \& Mendes de Oliveira
(2005) applied a similar technique
to search diffuse light in compact groups of galaxies.

This method involves a thresholding of the data in the wavelet space
in order to remove the noise locally without smoothing the
astrophysical signal.  
Wavelet coefficients are considered to be
significant at a given scale when their absolute magnitude exceeds
three times the RMS fluctuation expected for the wavelet coefficients
of white noise at the same scale.  The image analysis was performed
twice. From the first subset of significant wavelet coefficients, a
positive image in real space is restored. This image is subtracted
from the original one and the wavelet transform of the residual is
thresholded with thresholds similar to those in the first
iteration. 
Applying the restoration algorithm to the second subset of
coefficients enables one to obtain an image including the previously
hidden significant features.

In this way, adding together the two restored images with a 
constraint for the result to be positive (cf. the subtraction step) 
enables one to
obtain a more adequate map of the small scale structures within the
image. This last map can then be subtracted from the original image,
leading to a new residual image where large scale sources
can be quite easily searched for. 
Hence, the final products of this
process are an image of what we call the objects, i.e. the signal with
a characteristic scale smaller than the maximal scale used, and a
residual image exhibiting mostly features with a characteristic scale
larger than this maximum value. This value is $\sim$7 arcsec for the first
iteration and $\sim 2$ arcsec for the second iteration.

\begin{figure}
\centering
\psfig{file=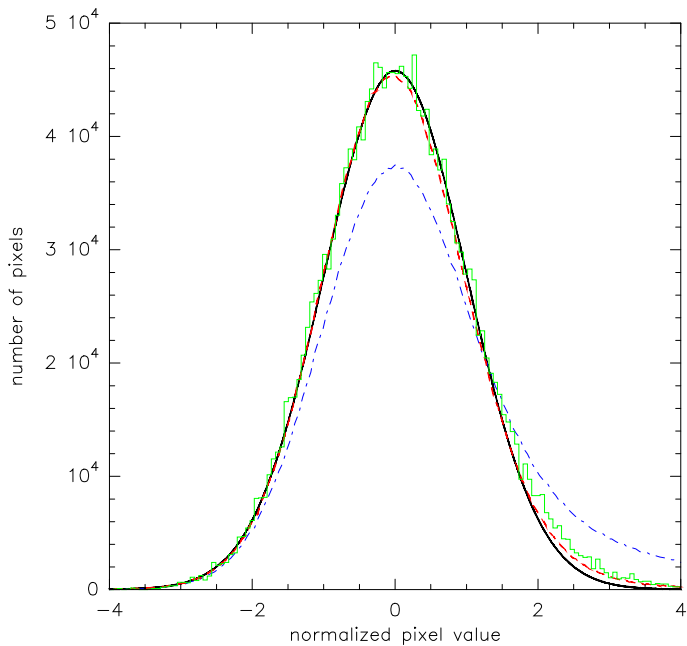,width=0.50\textwidth}
\caption{Comparison between the distibution of pixel values before and
after the removal of small-scale features in the WFPC2 F814W image.
The pixel values are normalized with respect to the rms fluctuation
measured in the residual image.  The blue dot-dashed line and the red
dashed line show the distributions of pixel values for the initial and
the residual images, respectively.  The black solid line stands for a
Gaussian fit of the pixel distribution in the residual image.  The
green histogram shows the distribution of pixel values in the residual
image for an empty background region (with size 200 $\times$ 200
pixels).  }
\label{noise}
\end{figure}

We checked that the noise in the residual image has nearly gaussian
statistical properties, as shown in Fig.~\ref{noise}. The difference between 
the pixel histogram and the gaussian fit results in slightly decreasing the 
statistical significance
of the detected features (a 3$\sigma$ level is, e.g., 
translated into a 2.5$\sigma$ level). First, the
applied procedure succeeds in removing the tail of the distribution
related to bright pixels linked to objects (i.e., stars and
galaxies). Second, the histogram of pixel values in the residual image
is nearly gaussian. The pixel value distribution of the non drizzled
residual image shows the same behaviour. Therefore, the correlation
between neighbouring pixels introduced by drizzling has a very small
effect and we have neglected it when computing the variance of the
noise. Note that the statistical significance of any feature detected
in the residual image has been estimated with respect to the gaussian
fit.

A few elongated
objects which appear to be possible tidal debris within the cluster
are not detected as diffuse sources by the present approach because
they have characteristic sizes smaller than 2 arcsec.  
Finally, we have checked that known strong gravitational lensing
features are not misclassified as diffuse sources or tidal debris, by
verifying that they are all included in the object map in all filters.
The results of the described multi-scale method 
are described in Sect.~\ref{images}.

\subsection{Serna \& Gerbal (1996) hierarchical analysis}
\label{spectro}

The redshift  
catalogue is presented in Covone et al. (2006a). 
This is a complete catalog down to total magnitudes of V$_{606}=21.5$
over the central $\sim 1 \, {\rm arcmin}^2$. 
We selected galaxies between $z=0.220$ and $z=0.244$ in order to remove
obvious interlopers from the redshift histogram, 
and applied to this sample the
Serna \& Gerbal (1996, hereafter SG) hierarchical method
with the aim of detecting substructures within the cluster core.

This method calculates the relative binding energies of all the 
galaxies on the basis of the measured redshifts and magnitudes. 
We used a mass-to-light (M/L) ratio in the B band of 450 
(see, e.g., \L okas \& Mamon 2003). 
We checked that using M/L between 320 and 580 
(i.e., large enough to encompass most of the M/L ratios in galaxy
clusters) does not
change, both qualitatively and quantitatively, the results. 

We identify four groups of galaxies:

 - a first group of four galaxies (hereafter, G1)
 in the central region which
has the largest (in absolute value) binding energy within the 
selected sample;

 - a pair of galaxies (G2) south west of G1 having a
binding energy $\simeq 6$ times lower than that of G1;

 - two minor galaxy pairs north of G1 (hereafter, G3) and south of
G1 (hereafter, G4), with respective binding energies 40 and 1200 times
smaller than that of G1.

\begin{figure}
\centering
\psfig{figure=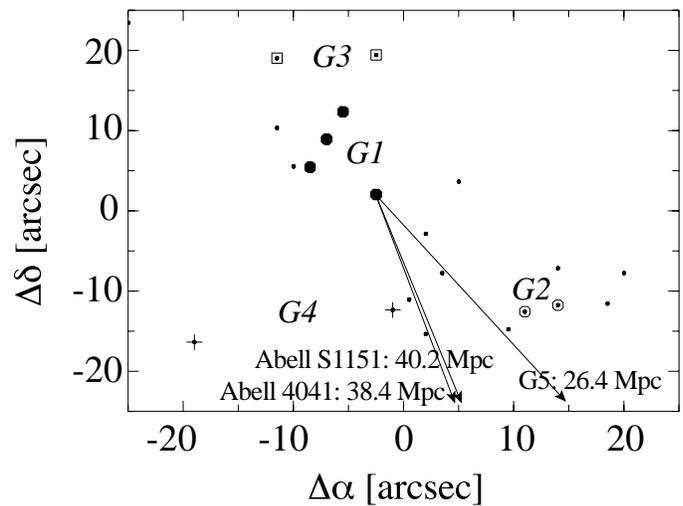,width=0.50\textwidth,angle=0}
\caption{Location of the subgroups detected within Abell~2667.  Dots:
positions of galaxies with redshifts between z=0.22 and z=0.244 in the
Abell~2667 IFU field of view (in arcsec relative to the geometrical
center of the galaxies measured spectroscopically). The galaxies
forming the central group according to the Serna \& Gerbal method are
indicated with filled black circles.  The open circles show two
galaxies forming the G2 pair towards the south west of the
cluster. Squares are for the G3 pair and stars for the minor G4 pair.
Lines indicate the directions of major nearby structures (Abell~4041,
Abell~S1151 and G5). North is up and East to the left.}
\label{fig:groups}
\end{figure}

In view of these numbers, we will consider in the following only 
the groups G1, G2 and G3. 
The positions of the galaxies in these structures are shown
in Fig.~\ref{fig:groups}. 
G2 and G3 have mean redshifts of $z= 0.2331$ and 0.2262, respectively,
therefore they are galaxy pairs within Abell~2667.

\section{Diffuse light in \a26}
\label{images}

\subsection{Imaging of the diffuse sources}

\begin{figure*}
\centering
\qquad\qquad\qquad\qquad\qquad\qquad 
\mbox{
\kern -0.2cm
\kern -0.1cm  
\kern -0.1cm  
}
\caption{Residual WFPC2 images in the F450W (left), F606W (middle) and
F814W (right) bands. Each figure is
1.2$\times$1.2~arcmin$^2$.  North is up and East to the left.}
\label{fig:wave}
\end{figure*}

Using the multi-scale approach described in Sect.2, we have identified
two distinct diffuse sources in the central $200 \times 200 \, {\rm
kpc}^2$.  The brightest source (DS1, hereafter) is located about 20
arcsec South-West of the cluster center, while the second one (DS2,
hereafter) is about 25 arcsec from the cluster center in the
North-East direction.
As shown in Fig.~\ref{fig:wave}, diffuse emissions are appearing at the
same positions and with roughly the same extensions in the three
residual images.

In order to estimate the statistical significance of the detections of
the two diffuse sources, we estimated the mean and standard deviation
of the images in the three bands in an external area where no diffuse
source was present. We compared this value to the mean level of the
two diffuse sources. This allowed us to estimate the values of the
detection significances given in the last columns of Table~\ref{table:DS}. 

The DS1 source is detected in the residual images of the two redder
filters at respective peak significance levels of 1.9$\sigma$ and
2.7$\sigma$ in the F606W and F814W filters, respectively. 
Its average surface
brightness is V$_{606} \simeq 26.2$ mag arcsec$^{-2}$. 
There is also a possible detection for DS1 in the F450W filter but with a 
lower significance ($\sim 1 \sigma$); 
this detection corresponds to a total
magnitude of  B$_{\rm 450}=21.9$.

The DS2 source is detected with lower significance levels of 1.5
 $\sigma$ and 2.1 $\sigma$ in the F606W and F814W images,
 respectively.  Its average surface brightness is V$_{606} \simeq
 26.7$ mag arcsec$^{-2}$.
Table~\ref{table:DS} summarizes the photometric properties of these
two diffuse sources. Total magnitudes in the three bands were computed
in the areas where the sources were detected in the F814W filter.

A small compact and well defined object is present inside the
diffuse source DS1 and will be excluded from the rest of our analysis. 
A second very faint object is also observed within DS1 (see
Fig.~\ref{fig:CDS}) and will be called the ``compact diffuse source''
(ComDif). Photometry and morphological characteristics for this object 
have been computed using SExtractor.

\begin{figure}
\centering
\caption{Zoomed image (F814W filter)
of the region around the ComDif diffuse source (at the center of the image).
Contour levels correspond to surface brightness levels  of 
I$_{814}$ = 23.0, 22.0, 21.0 mag/arcsec$^2$.
The field is 0.5$\times$0.3~arcmin$^2$ wide. 
North is up and East to the left.
The elongated object $\sim 10$ arcsec W of the center
is a background lensed object (Rc in Covone et al. (2004);  
photometric redshift $z=1.15$).}
\label{fig:CDS}
\end{figure}

\begin{figure}
\centering
\psfig{file=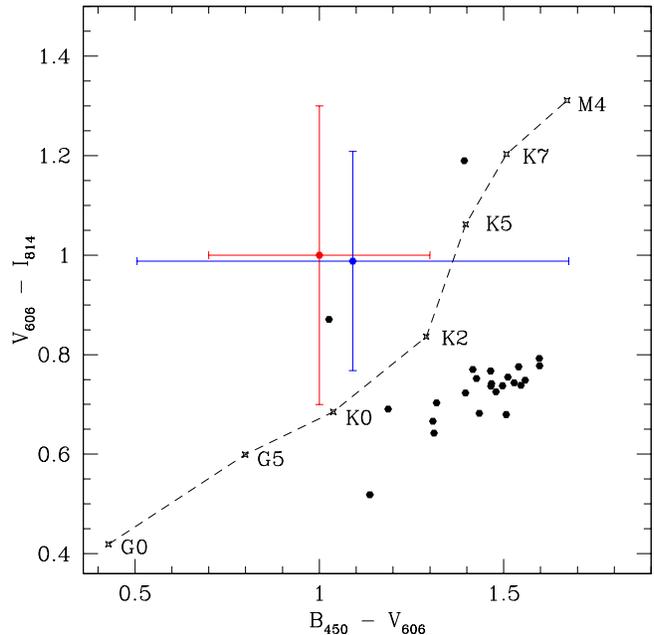,width=0.50\textwidth}
\caption{Color-color diagram for the spectroscopically 
confirmed cluster galaxies (filled circles), 
the DS1 (red symbol with smaller error bar along the B-V axis)
and ComDif (blue symbol with smaller error bar along the V-I axis) sources and 
the Pickles (1998) 
late-type star templates redshifted at $z=0.23$ 
(crosses connected by a line).}
\label{fig:color}
\end{figure}

We have compared the colors of the diffuse sources to those of the
spectroscopically confirmed cluster galaxies 
and those of late-type stars (from the Pickles 1998 library)
redshifted at $z=0.233$ (see Fig.~\ref{fig:color}). 
The two sources have similar colours, within the present photometric
uncertainities, 
and close to those of evolved and K-type
giants, as already found for the
tidal debris in Centaurus A by Calc\'aneo-Rold\'an et al. (2000).
They appear to have average V-I
colors redder than the evolved cluster population and bluer
average B-V colors.

When compared to the colors of the redshifted spectral
templates of galaxies from the Kinney-Calzetti spectral atlas 
(Kinney et al. 1996) -- but without taking into account evolution --  
we find that their B-V color is 
between that of an Sbc galaxy and a S0 galaxy.

\subsection{Spectra of DS1 and ComDif}

\begin{figure}
\centering
\caption{Positions of the fibers of the VIMOS-IFU considered to obtain
the spectrum of the diffuse emission discovered SW of the
cluster (DS1).  
The hole in the grid corresponds to an obvious object that
was taken out. North is up and East to the left.}
\label{fig:im_ifu}
\end{figure}

Using the spatial information of the detected diffuse sources, we
extracted their 
spectra from the VIMOS-IFU data cube 
(Covone et al. 2006a), avoiding the known compact sources
in that area (Fig.~\ref{fig:im_ifu}).  
In total, we used 33 and 267 fibers (i.e., about 11 and 91 arcsec$^2$)
for ComDif and DS1, respectively. 
We could not obtain a reliable spectrum for  
the lower surface brightness source DS2 as it 
was located in a region covered by a set of dead fibers.

\begin{figure}
\centering
\psfig{file=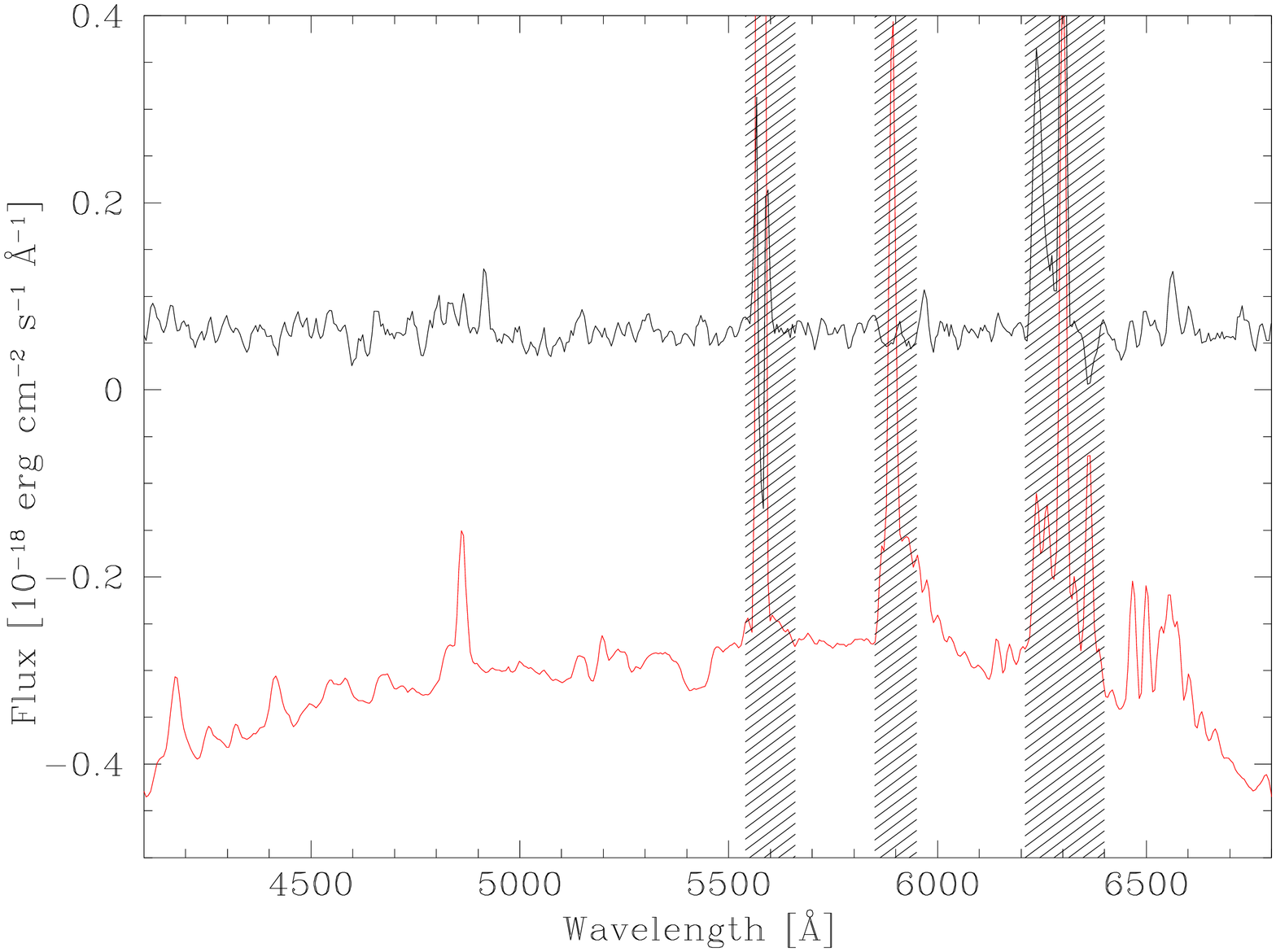,width=0.50\textwidth,angle=270}
\psfig{file=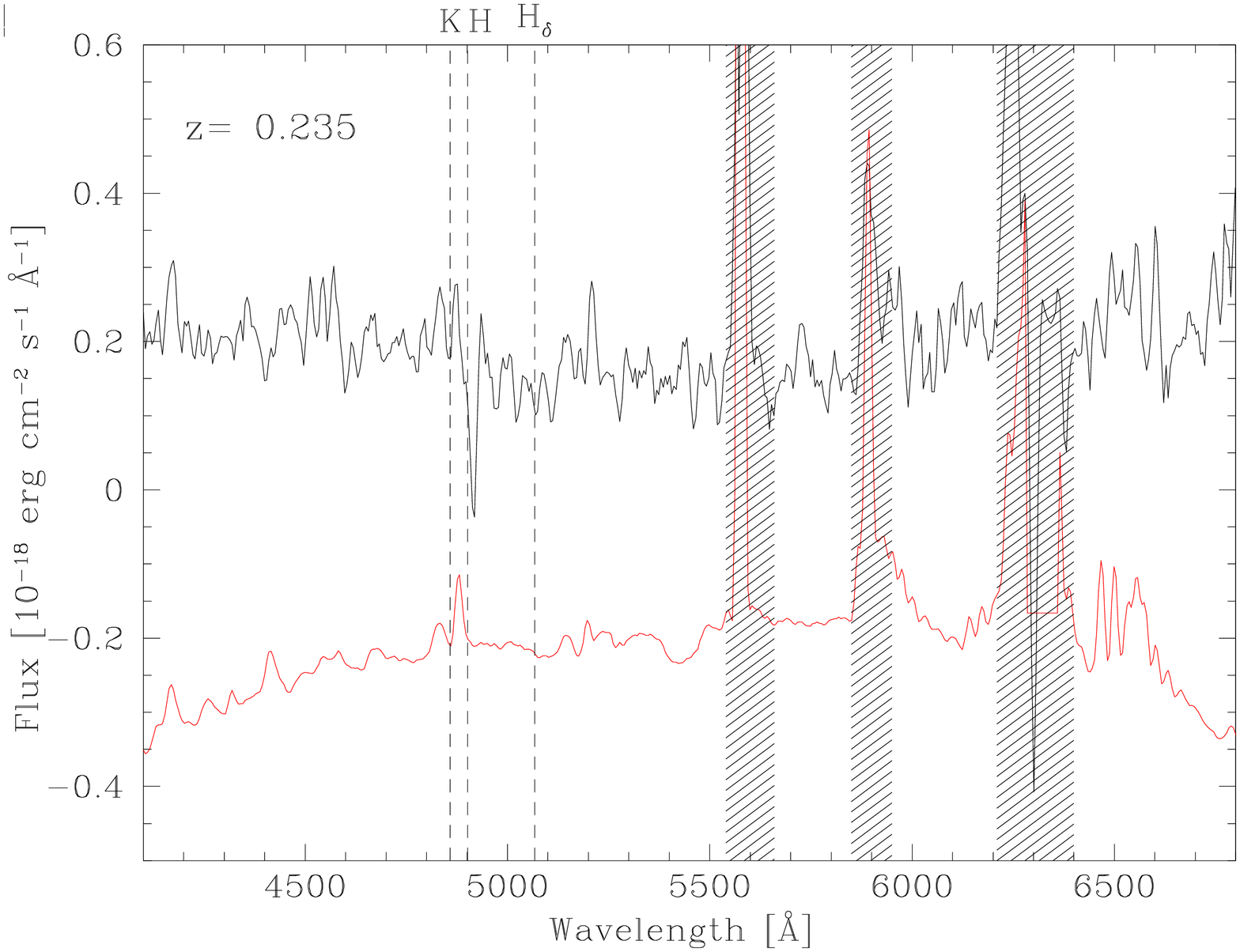,width=0.50\textwidth,angle=270}
\psfig{file=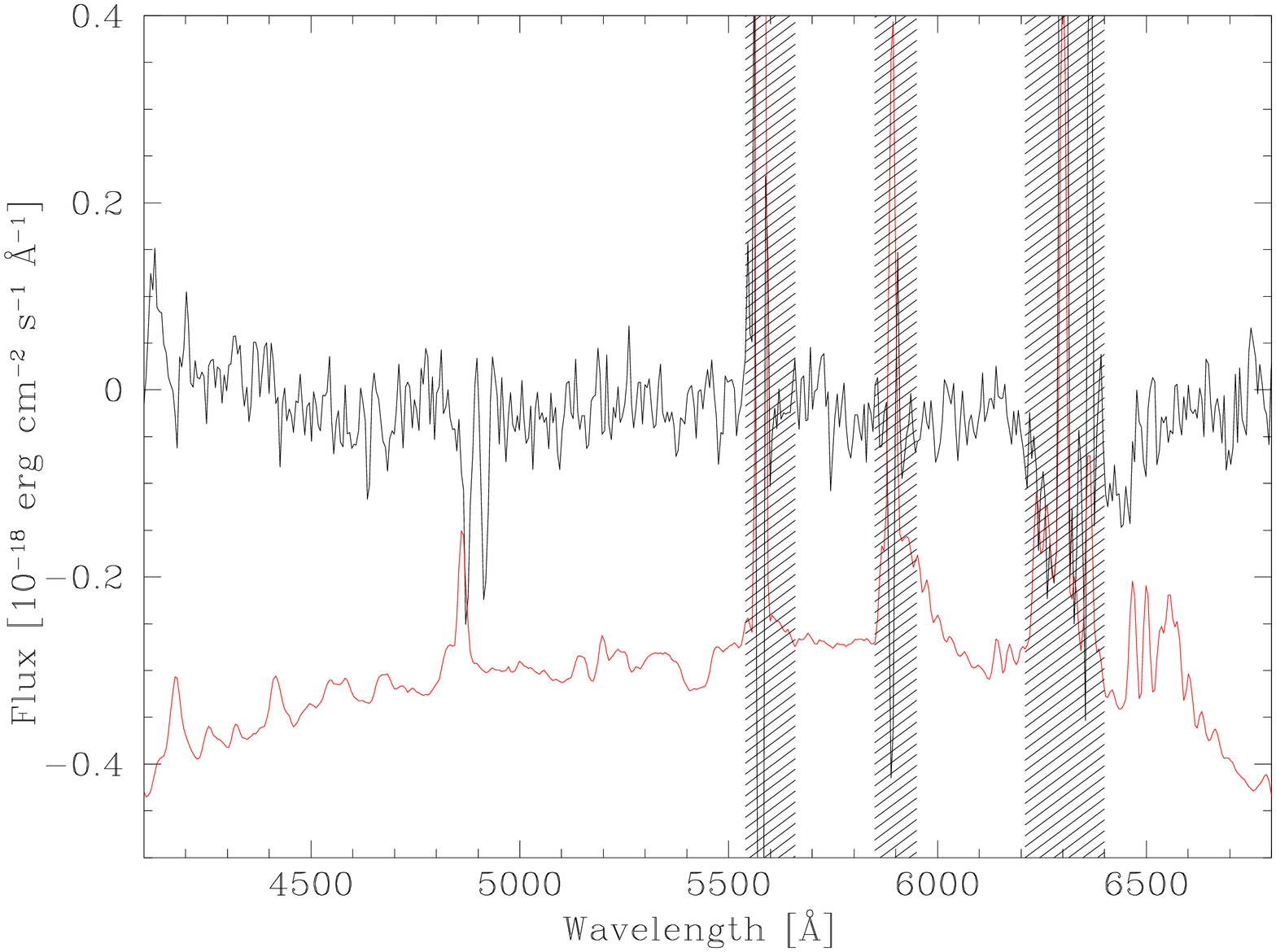,width=0.50\textwidth,angle=270}
\caption{Spectra of the entire DS1 diffuse source (upper panel), 
of the compact diffuse source ComDif (middle panel), 
and of an empty region nearby (lower panel),
together with
a scaled background signal at the same locations (in red).
Spectra have been Gaussian smoothed with a FWHM=15 \AA.
The expected positions of some absorption lines
at redshift $z=0.235$ are shown (see text for discussion).
Shaded regions mark the spectral ranges where sky signal is dominant.}
\label{fig:spectres}
\end{figure}

We plot the spatially averaged, 
one-dimensional spectra of the DS1 and ComDif regions in
Fig.~\ref{fig:spectres}, together with a local sky spectrum (i.e., the
summed spectrum from nearby fibers in the same quadrant, not covering
any detected object).

The detection of the spectrum from faint and extended sources 
in a VIMOS-IFU data-cube is a demanding task.
The large extension of the source makes the removal of the background
signal very difficult.  In this situation the statistical approach
generally used to remove the background signal in the VIMOS-IFU
(Zanichelli et al. 2005), i.e. computing the sky in each module of 400 fibers, is clearly not
appropriate, since it would remove any signal from extended sources
which may cover a large fraction of the considered fibers.
Therefore, we produced a data-cube with no such background removal,
and evaluated the sky in regions with no detected objects close to the
considered sources.  In order to take into account the fact that
fibers have different transmissions, fiber spectra have been
renormalized using the flux measured in a given sky line, as described
in Zanichelli et al. (2005; see also Covone et al.  2006b).
However, as the background signal in the VIMOS-IFU is a strong
function of position, this approach introduces further noise in the
extracted spectra.  
Unfortunately, this is particularly true in the spectral range
$\lambda \sim 4800 - 4900$ \AA \, where the signal from the CaII H and
K absorption
lines from objects at rest in the cluster potential well is expected.

As shown in Fig.~\ref{fig:spectres}, we clearly detect the continuum
of both ComDif and DS1, thus supporting the photometric detection
since no such continuum is detected in areas without visible ICL.

However, the low S/N ratio and the noise spikes due to the spatially 
varying background prevent us
from detecting any significant absorption feature.
The redshift of ComDif would be 0.235$\pm$0.003 if we assume the
H$\&$K lines as real. Deeper data would be needed to confirm these
results.

We note that no emission line is detected at the given positions. 
We can set an upper limit
($f_{\rm l} \simeq  0.7 \times 10^{-18} \, {\rm erg \, s}^{-1} \, {\rm cm}^{-2}$) 
to any emission line 
(covering a spatial extension of one fiber). 
Such an emission line absence most probably implies that DS1 is made of
stars and not of ionized gas.

\section{X-ray analysis}
\label{xrays}

We have used both \textit{Chandra} and XMM-\textit{Newton} archival data. We
have taken advantage of the \textit{Chandra} high resolution imaging to search
for fine morphological structures in the gas distribution and of the
XMM-\textit{Newton} high sensitivity to obtain spatially resolved X-ray
spectra.

\subsection{Chandra data}

\begin{figure}[!t]
\centering
\caption[]{\textit{Chandra} ACIS-S3 X-ray adaptively smoothed image of
Abell~2667 in the [0.3--8.0~keV] band, corrected by the exposure map
and binned by a factor of 4 (1 pixel = 2~arcsec).  The
HST/WFPC2 contours (filter F814W) are shown as black lines.
\label{fig:smooth}}
\end{figure}

\begin{figure}[!t]
\centering
\caption[]{XMM temperature map of the X-ray gas in the central
region. The contours correspond to the \textit{Chandra} surface
brightness as in Fig.~\ref{fig:smooth}.  \label{fig:kTmapZoom}}
\end{figure}

The \textit{Chandra} observation was performed in June 2001 with the ACIS-S
detector (observation ID 2214, P.I. S. Allen). The data was taken in Very
Faint mode with a time resolution of 3.24~sec, during 11~ks. The CCD
temperature was $-120^{\circ}$C.

We have re-run the ``pipeline'', following the standard data reduction
procedure,
Standard Data Processing, producing new level 1 and 2 event files with CIAO
version 3.2 and CALDB version
3.0.0\footnote{\texttt{http://asc.harvard.edu/ciao/}}. We have restricted our
data reduction and analysis to the back-illuminated chip, ACIS-S3 (the cluster
is completely contained in this CCD chip).

We have used the ``blank-sky'' CTI-corrected ACIS background event files, made
available by the ACIS calibration
team\footnote{\texttt{http://cxc.harvard.edu/cal/Acis/WWWacis\_cal.html}}. The
background events were filtered, keeping the same grades as the source events,
and then were reprojected to match the sky coordinates of the Abell~2667 ACIS
observation.

We restricted our analysis to the range [0.3--8.0~keV], because above 8.0~keV
the X-ray emission is largely background-dominated.

\subsection{XMM-Newton data}

Abell~2667 has been observed in June 2003 (revolution 647, exposure 0148990101,
P.I. S. Allen) during 26~ks, in standard Full Frame mode using the ``medium''
filter with the EPIC MOS1, MOS2 and PN detectors.

We have used the standard procedure and produced ``clean'' event files. The
light-curves in the [10--12 keV] band showed that there were flaring events
during the observation. Filtering out the periods with flares reduced the
exposure times to 22,226~s, 22,502~s, and 15,092~s for MOS1, MOS2 and PN,
respectively.

With the cleaned and filtered event files, we have created the redistribution
matrix file (RMF) and ancillary response file (ARF) with the
XMM-Newton Science Analysis System (SAS) tasks
\texttt{rmfgen} and \texttt{arfgen} for each camera and for each region that
we have analyzed.

The background was taken into account by extracting spectra from the blank sky
templates described by Lumb et al. (2002), and reprojected to the coordinates
and roll angle of Abell~2667.

\subsection{\textit{Chandra} X-ray imaging}

We have constructed an adaptively smoothed image in the [0.3--8.0~keV] band
using the \textsc{csmooth} tool from CIAO. The ACIS-S3 point spread
function (PSF) has an on-axis spatial FWHM $\approx 0.8$~arcsec, weakly dependent on
the energy, and the unbinned pixel size is $0.5$~arcsec\footnote{
\texttt{http://cxc.harvard.edu/cal/Acis/Cal\_prods/psf/
psf.html}}. 
The exposure map was generated by the script \textsc{merge\_all}, where we 
have calculated the spectral weights, needed for the instrument map, using the
cluster total spectrum, i.e., the spectrum obtained inside a circle of radius
of 1~arcmin centered on the cluster center.

Fig.~\ref{fig:smooth} shows the \textit{Chandra} ACIS-S3 image with HST
isocontours superimposed. The offset between the center of the main
X-ray emission and the central galaxy is only about 1.0~arcsec (i.e., 
less than $4 \, h_{70}^{-1}\,$kpc). The
overall X-ray emission is elliptical and aligned with the galaxy
distribution. A secondary X-ray peak is observed about 10~arcsec
South West of the cluster center, 
at the position of a bright galaxy. 
Further along the same direction, $\simeq 29$~arcsec away from
the center, there is a third, smaller X-ray peak coinciding with a
galaxy pair.

\subsection{X-ray spectral analysis and temperature map}

The EPIC-MOS and PN PSF have a FWHM $\approx 5$~arcsec, but
their PSF have somewhat extended wings and the half energy width is
$\approx 14$~arcsec\footnote{
\texttt{http://xmm.vilspa.esa.es/external/
xmm\_user\_support/documentation/uhb/index.html}} The spectral
analysis is not limited by the spatial resolution but rather by the
binning necessary to achieve the sufficient signal-to-noise ratio
required for spectral fitting.

The temperature map was computed from XMM-Newton data following Durret et al.
(2005). Briefly, the event files are rebinned with a pixel size of $25.6$
arcsec and for each pixel in the grid we compute the RMF and ARF matrix and
fit a \textsc{mekal} (Kaastra \& Mewe 1993; Liedahl et al. 1995)
 plasma model using \textsc{xspec} version 11.2.

For the inner region only, a rectangle of $76.8 \times 128$ arcsec$^2$, we
used a higher resolution pixel of 12.8~arcsec. We set a minimum count number
of 1200 counts per pixel.

When we had enough counts, the spectral fit was done with the hydrogen column
density fixed at the local Galactic value. 
For each pixel, we estimate
$N_{\rm H}$ using the task \texttt{nh} from \textsc{ftools} (which is an
interpolation from the Dickey et al. 1990 Galactic $N_{\rm H}$ table).

Fig.~\ref{fig:kTmapZoom} shows the temperature map in the central region of
Abell~2667. The gas is cooler in the central $R \approx 13$ arcsec, with a
temperature $kT = 4.2 \pm 0.3\,$keV. The coolest pixel coincides with the main
X-ray intensity peak. At the limits of the map, the temperature reaches about
12~keV. 
The ratio between the high and low temperatures is about 3, as
observed in most cooling-core clusters (e.g. Piffaretti et al. 2005).

Contrary to the surface brightness map, which shows elliptical symmetry, the
temperature map shows a cooler lane, roughly stretching from North to South,
with hotter regions towards North East, South West and, to a smaller level,
North West relatively to the cluster center. 
The general direction of the cold lane is approximately
perpendicular to the principal axis of the X-ray emission.

\begin{figure}[!h]
\centering 
\includegraphics[width=\columnwidth]{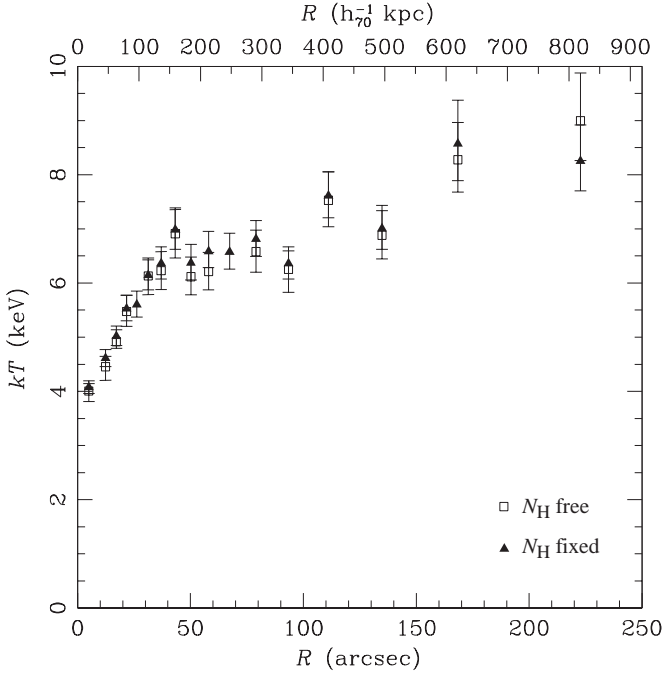}
\caption[]{XMM temperature profile obtained in concentric circles,
with the hydrogen absorption column either fixed to the Galactic
value (black triangles) or free to vary (empty squares). The
temperature is in keV and the radius in arcsec (bottom axis) or in
kpc (top axis). \label{fig:kTprofile}}
\end{figure}

We drew the temperature profile by extracting spectra in concentric
circles (see Fig.~\ref{fig:kTprofile}). This profile confirms that the
temperature is notably lower in the innermost 150~kpc of the cluster,
remains roughly constant up to a radius of $\sim$350 kpc, then
increases notably in the regions where the temperature map has indeed
revealed the presence of hotter zones (Fig.~\ref{fig:kTmapZoom}). 
Note however that the error bars on the temperature are large in the
outermost bin.

\begin{figure}[!h]
\centering
\includegraphics[width=\columnwidth]{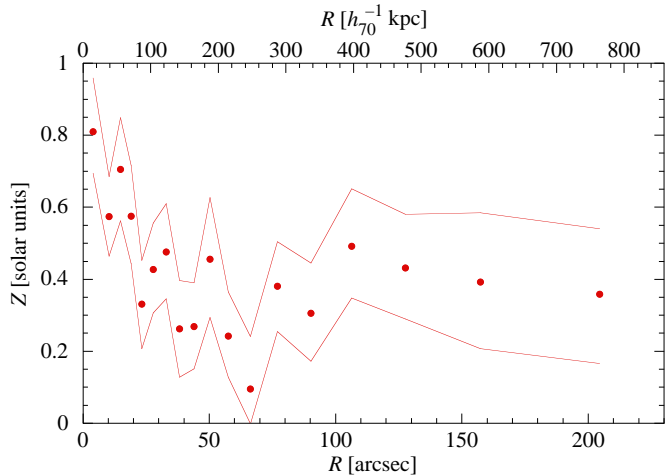}
\caption[]{XMM metallicity profile of the X-ray gas in concentric
circles, in solar units.  Continuous lines correspond to the 90\% confidence
level interval (i.e., 2.7 $\sigma$).}
\label{fig:Zmap}
\end{figure}

In a similar way, fitting a single \textsc{mekal} model, we drew the metallicity
profile of the X-ray emitting gas, shown in Fig.~\ref{fig:Zmap}. It shows high
values between 0.6 and 0.8 (in solar units) in the innermost 100~kpc, then
varies between 0.2 and 0.5 at larger radii. The mean emission-weighted
metallicity within 200 arcsec is about $0.4 \, Z_{\odot}$, 
a value usually found in nearby rich clusters (e.g., Fukazawa et al. 2000). 
Fixing or not the hydrogen column density only has a
very small effect; therefore the results related to the metallicity presented
below are always obtained letting $N_{\rm H}$ free.

\begin{figure*}[!h]
\centering
\caption[]{Spectral extraction regions are shown superimposed on the
XMM (left) and HST (right) images. The yellow circles (12
arcsec radius each) show the regions corresponding to the diffuse
light sources DS1 and DS2.  The inner white circle (14 arcsec
radius) is excluded and the outer white circle (38 arcsec radius)
binds the ring used as a control.  \label{fig:DS1DS2regs}}
\end{figure*}

Finally, we have obtained the temperature and metallicity at the positions of the two
diffuse sources, DS1 and DS2. Fig.~\ref{fig:DS1DS2regs} shows the regions used
for spectral extraction, as well as a control ring that excludes the DS2
region and part of the DS1 region. 

The temperature and metallicity in the DS1 region are $kT = 4.8 \pm 0.3\,$keV
and $Z = 0.50^{+0.11}_{-0.10}\, Z_{\odot}$ (90\% confidence level),
respectively. For the DS2 source we obtain $kT = 6.1^{+0.8}_{-0.6}\,$keV and
$Z = 0.56^{+0.19}_{-0.18} Z_{\odot}$.
In order to check if the metallicity in DS2,
the region with the higher metallicity and farther from the center, is higher
than in its neighborhood, we extract a spectrum in the control ring around the
position of the DS2 source, but excluding it and part of the DS1 source (see
Fig.~\ref{fig:DS1DS2regs}).

\begin{figure}[!h]
\centering
\includegraphics[width=\columnwidth]{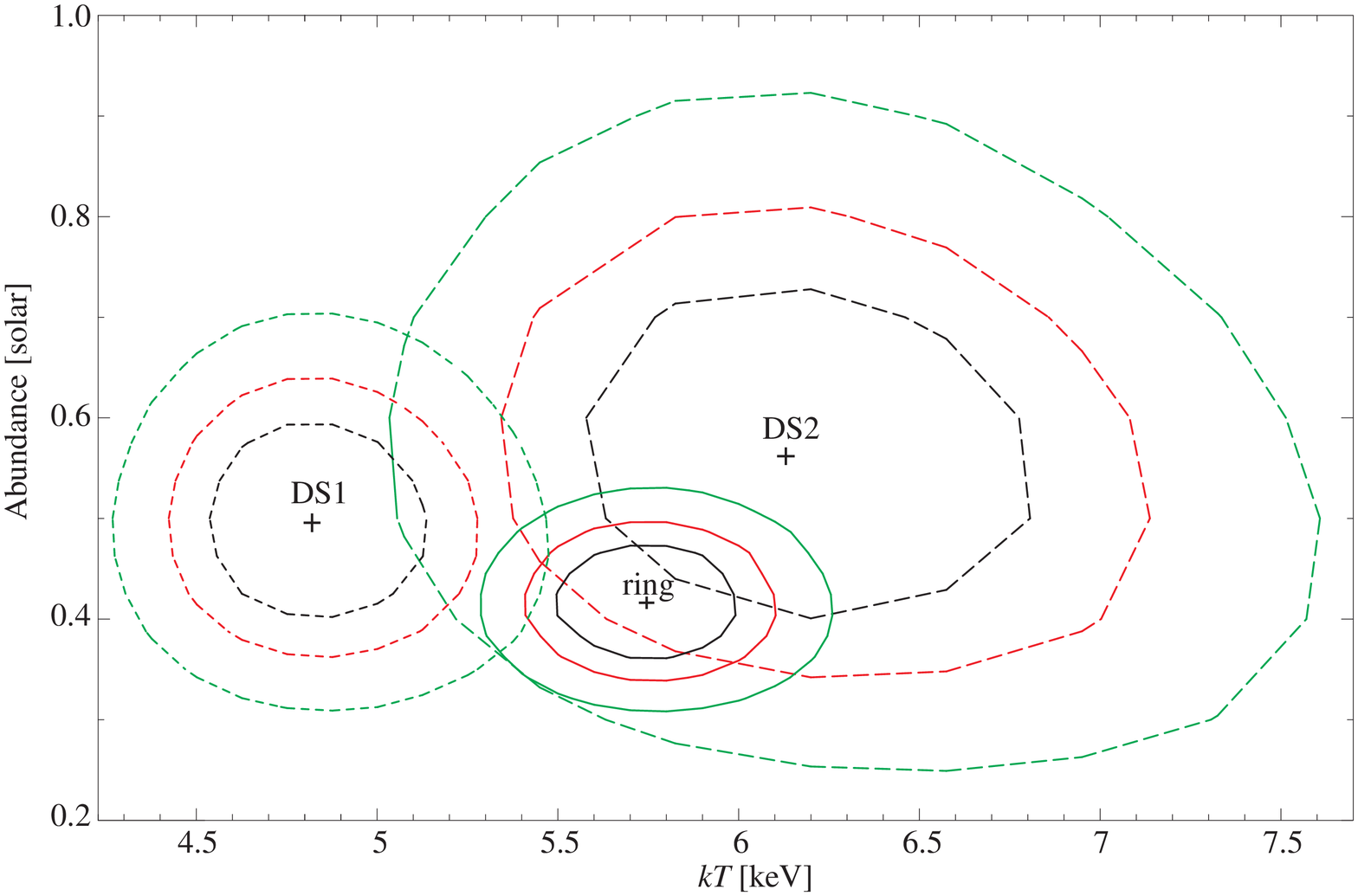}
\caption[]{Confidence levels (1, 2 and 3 $\sigma$, in black, red and green,
respectively) for the spectral fits of regions DS1 (short dashes), DS2
(long dashes) and the ring excluding both regions (full lines) (see text for
details). 
\label{fig:contlevel}}
\end{figure}

We have plotted the confidence level curves for the quantities
$kT$ and $Z$ for DS1, DS2 and the control ring in
Fig.~\ref{fig:contlevel}. The ICM temperature and metallicity of the
control ring are $kT = 5.7 \pm 0.3\,$keV and $Z = 0.42 \pm 0.06\,
Z_{\odot}$. Although the metallicity of the regions associated with DS1 and
DS2 may be higher than that in their surroundings, the
difference is significant only at $\sim 1 \sigma$.
Therefore, a deeper X-ray exposure is required to confirm the
differences in metallicities found with the present data.

\section{The building history of Abell~2667}
\label{discussion}

\subsection{Diffuse light}

By applying a multi-scale wavelet analysis and reconstruction
technique to the HST images of the cluster Abell~2667 in three bands,
we have shown the probable existence of one large scale source
of diffuse emission (DS1), in which is embedded a faint and more
compact extended source (ComDif), likely at the same redshift as the
cluster.  A second diffuse source may be present (DS2) at a lower
significance level. 

We have measured the optical colors of these sources.
The colors of the two sources DS1 and ComDif are close to that of
K-type stars at the cluster redshift (Fig. 5), 
implying a population of evolved stars. 
These colors are slightly bluer than
those found in a simulated Coma-like cluster at $z=0$ 
by Sommer-Larsen et al. (2005), 
probably because of the younger age of the detected diffuse light systems 
(at least 3.7 Gyrs younger since located at $z$=0.233).

VIMOS-IFU spectroscopy
reveals a continuum signal at the location of the sources
DS1 and ComDif, with no apparent emission line
brighter than a flux limit
$f_{\rm l} \simeq 0.7 \times 10^{-18} \, {\rm erg \, s}^{-1} \, {\rm cm}^{-2}$.
The identification of the Ca H and K absorption lines
(expected in an evolved stellar population)
at the cluster redshift is problematic because of the strongly varying
background at the interested spectral position.

The X-ray gas in the DS1 and DS2 regions may be
metal-richer than their immediate neighbourhood 
(Fig.~\ref{fig:contlevel}) but this is only a 1$\sigma$ difference. 
While deeper X-ray data are necessary to confirm these findings,
we note that
Zaritsky et al. (2002) have shown that the ICL population can significantly 
contribute to the metal enrichment of the ICM, via 
supernova winds, 
an {\em in situ} mechanism which also
suggests a spatial correlation between
metal richer regions and ICL sources. 

Recent numerical simulations also have not yet addressed the question
of a possible spatial correlation between the ICM metallicity and the
presence of ICL sources.  Globally, the metallicity profile in
Abell~2667 (see Fig. \ref{fig:DS1DS2regs}) is in good agreement with
simulations by Romeo et al. (2005).
The values close to half solar are also in good agreement with Romeo
et al. (2005).

The quantity of matter included in the two detected diffuse light
sources remains modest compared to the total cluster mass.  While the
large scale diffuse light contribution to the Coma cluster was close
to the luminosity of a cD galaxy (Adami et al. 2005), here this
diffuse light only represents a small fraction (about 10-15\%) of the
luminosity of the main galaxy, which is I$_{814}=17.2$ (absolute
magnitude $-23.1$).  This is in good agreement with the work of Lin
$\&$ Mohr (2004) showing that the ICL contribution increases with
cluster mass.We note that we have detected only the sources of diffuse
light larger than $\sim$2 arcsec and with a surface brightness
brighter than $\sim$26 in the F814W filter.  Therefore, our total
light fraction estimate is only valid for large scale sources.

The total contribution of these diffuse light sources is equivalent to an
I$_{814}=19.4$ galaxy (absolute magnitude $-20.9$). If we assume that
diffuse light is proportional to the infall activity onto the cluster (as
suggested by Willman et al. 2004), Abell~2667 is, therefore,
apparently a more isolated cluster than the Coma cluster in the cosmic
web. However, the link between infalling groups and diffuse light sources is
far from clear and this has to be confirmed by a more specific study of the
environment of these two clusters.

Finally, we note that this fraction is in agreement with the findings of
Zibetti et al. (2005) who summed the faint signal from a large sample
of rich clusters from the SDSS
at roughly the same cosmic epoch, and with the numerical simulations by
Willman et al. (2004), who have followed the time evolution of the
ICL fraction in a Coma-like cluster.

\subsection{Diffuse light sources as part of the brightest cluster galaxy envelope?}

In this work we have made no strict distinction between the brightest cluster galaxy envelope and
the diffuse sources
found by using the multiscale approach.

Indeed, while in numerical simulations it is possible to separate the
two stellar populations on dynamical grounds, this is not possible
using photometric data.
Kelson et al. (2002) have used deep long-slit spectroscopic data to
show that, in the galaxy cluster Abell~2199, the outskirts of the
brightest cluster galaxy halo share the same velocity dispersion than
the cluster galaxies, with a monotonic increase from a few kpc from
the brightest cluster galaxy center outwards.  This behaviour thus
supports a view in which ICL sources and the brightest cluster galaxy
envelope originate from the same physical processes.

The detected diffuse sources could however be considered
as being part of the brightest cluster galaxy halo in terms of the usual photometric
definition.  In other words, an alternative explanation to the
presence of sources of diffuse light would be that these are not
independent from the dominant galaxy envelope.  If we fit elliptical
isophotes on the F814W HST image after excluding the other visible
galaxies, using the {\tt ellipse} package in STSDAS, we find that the
surface brightness profile starting from the dominant galaxy center
approximately follows an r$^{1/4}$ law with an ellipticity growing
from $\sim$0.2, close to the center, to $\sim$0.6, 30 arcsec from the
center.

In this picture, the sources DS1 and DS2 do not appear as major residuals
over the r$^{1/4}$ law. This means that the sum of the possible
diffuse light sources and of the dominant galaxy contribution is
correctly modeled by an r$^{1/4}$ law, as long as we allow the
ellipticity to grow regularly from the dominant galaxy center to the
image borders.

It is difficult,however, to constrain unambiguously the true ellipticity of the
dominant galaxy, i.e. the ellipticity corresponding to this galaxy and
not due to the addition of independent diffuse light sources. If the
true ellipticity of the dominant galaxy really increases radially up
to a value of 0.6 far from the center, then the detected diffuse
sources, even if they are real, can simply be the external parts of
the strongly elliptical central galaxy. If the true ellipticity of the
main galaxy remains constant, then the diffuse light sources are
independent entities. 

Note however that up to a certain point these two hypotheses are
similar, since it is difficult to estimate a sharp boundary between
the brightest cluster galaxy envelope and the nearby diffuse sources.
Accretions are probably occuring along the main cluster axis, which is
also that of the main central galaxy.  So, considering the diffuse
sources as independent entities that will soon be accreted by the
central galaxy in its external parts along the infalling direction, or
considering that the diffuse sources have already been accreted on to
the external parts of this galaxy (see, e.g., Richstone 1976) is not
very different.  A possible way to discriminate between these two
explanations would be to get sufficient S/N spectra of the diffuse
sources to compute their precise redshifts; unfortunately, our
spectroscopic data are not deep enough for this purpose.

Another caveat is that elongated $r^{1/4}$
halos are normally seen around ellipticals and they are in virial
equilibrium (see e.g. Arnaboldi et al. 2004, Mihos et al. 2005). So the
detected extensions could not be related to the direction of infalling material
at all. However, the good correlation between the brightest galaxy halo
extension and the infalling directions still suggests a relation
between these two components.

\subsection{Infalling directions}

A natural explanation would be that the diffuse sources are
concentrations of tidal debris and harassed matter expelled from
galaxies by tidal stripping, close galaxy interactions or strong tidal
forces close to the cluster center. However, there is only one clear 
distorted object (ComDif). The remaining of the diffuse sources appear quite
smooth. This hypothesis remains therefore to be tested with deeper images but
these sources could be the result of already well mixed old tidal tails.

These processes are expected to act on infalling galaxies and it
would not be surprising to find diffuse light sources along the main
infall directions, altough this view still needs support from
more observations and numerical simulations.

The X-ray analysis shows somewhat hotter regions south-west and
north-east of the cluster, which could be due to shock fronts
along these two directions produced by infalling material still
distant from the cluster center. The error bars on the X-ray gas
temperature in the outer zones are, however, large and this scenario has to be
consider with caution even if it would
agree with the general elongation of the X-ray gas emission 
(see Fig.~\ref{fig:smooth}) and with the
hypothesis of matter infalling along a north-east to south-west
direction.
This direction matches that joining DS1 and DS2. Moreover, the
dynamical analysis we performed using the SG method shows three
independent groups (G1, G2 and G3) again defining the same direction.
We suggest that G2 and G3 are infalling groups from the surrounding
cosmic web.

In order to find evidence for
this cosmic web, we used the 
NASA Extragalactic Database (NED) to search 
for known galaxy clusters in a $\sim$40 Mpc radius area around \a26\
and with redshifts between 0.21 and 0.26.  Only two rich galaxy
clusters have been found, Abell~4041 and Abell~S1151 located about
40 Mpc from Abell~2667.
These two clusters are located roughly in the same direction as
defined by the groups G2-G1 (see Fig.~\ref{fig:groups}).
We note, however, that the known cluster population at these redshifts
is far from being complete and we cannot  conclude that G2
is infalling onto the Abell~2667 cluster core coming from the
Abell~4041/AbellS~1151 complex, which is notably further away.  

We also searched for all known individual galaxy redshifts between 
$z=0.21$
and 0.26 in a large area around Abell~2667. A hundred redshifts are
available from the NED database (mainly from the 2dF survey)
and are concentrated in a strip extending from $\delta$=-28$^\circ$ to
$\delta$=-26$^\circ$. Applying the SG algorithm to these data, we
detected several bound structures, but only one with a binding energy
larger than that of G4 (of the order of that of G1) and sampled with
more than 10 redshifts. 
This structure is a group (hereafter G5),
located at $\alpha$=356.84$^\circ$ and $\delta$=-27.66$^\circ$, 
at $\simeq$ 6 arcmin from the Abell~2667   center.
Its redshift 0.221$\pm$0.001, and its
velocity dispersion 164 km s$^{-1}$, corresponding to a mass of 8 
$\times 10^{12} \, $M$_\odot$. 
This structure is located in the same direction as defined by groups G2-G1
(see Fig.~\ref{fig:groups}) and is located at about 26 Mpc from
Abell~2667.

The detections of G2, G5, Abell~S1151 and Abell~4041 along a direction
defined by the elongation of the X-ray emission and by the zones where
the X-ray gas is hotter, are all consistent with the hypothesis
of a filament crossing Abell~2667 and along which galaxies are flowing
onto the cluster, but of course in view of the limited S/N of all
our data this scenario remains speculative.

\subsection{G2 infalling characteristics}

G2 is perhaps directly responsible for the formation of the ICL source
DS1.  If this group is actually infalling onto the core of Abell~2667
we may expect that  the low-mass galaxies of
this group will be in advance along the infalling direction 
(see, e.g., Sarazin 1986).  As low
mass galaxies are easier to disrupt than massive ones, they could have
been destroyed before reaching the Abell~2667 cluster core, therefore
contributing to form the  sources DS1 and ComDif.  

G2 is constituted of two relatively bright galaxies (the fourth and
seventh brightest galaxies of the cluster, Covone et al. 2006a). 
This suggests that G2 already started to dissolve in the Abell~2667
potential, as such bright galaxies are expected to show an
associated  surrounding faint galaxy group. 
It is difficult to estimate the original mass of this group: 
the absence of any 
extended X-ray emission (see Fig.~\ref{fig:smooth}) acts in favor of a minor
group.  Recent numerical simulations by Tormen et al. (2004) show that
an infalling group with a relative mass compared to the main cluster
between 1 and 10$\%$ will keep its intra-group gas during less than 2
Gyrs, but will survive as an independent entity for a longer time
(about half of its dark matter staying clustered during up to 3 Gyrs). 
In this scenario, if
G2 already lost its gas but still exists as an independent pair,
this implies that it has been falling onto Abell~2667 for a time
between 2 and 3 Gyrs.
Assuming that the tidal gravitational forces from Abell~2667 start to
be effective in removing the group gas at roughly the virial radius,
the corresponding infall velocity of G2 would then be between 1000 and
1500 km s$^{-1}$.

\section{Summary}
\label{concl}

We have proposed a possible scenario for the building process in the core
of the galaxy cluster \a26\ based on the analyses of optical {\em HST}
imaging, {\em Chandra} and {\em XMM} X-ray data, and VIMOS-IFU
spectroscopy.  Here we summarize our main suggestions:

\begin{description}

\item[(i)] By means of an iterative multi-scale wavelet analysis and
reconstruction of {\it HST} images of Abell~2667 in three filters, we
have shown the presence of a zone of diffuse emission south west of
the cluster center (DS1); a second very faint object, the ``compact
diffuse source'' (ComDif), is observed within DS1.  These two sources
have similar colors, close to those of old, K-type stars at the
cluster redshift.  Another diffuse source (DS2) is also possibly
detected north east of the center in the reddest filter, but with
lower S/N.

\item[(ii)] Four dynamical groups are detected within the cluster core
using the Serna \& Gerbal (1996) hierachical method (G1, G2, G3 and
G4).  One additional group (G5) is identified much further away 
$\simeq 6'$ (1.34~Mpc) from the cluster center.

\item[(iii)] A clear spectroscopic continuum is detected for DS1 and
ComDif, though the strongly spatially varying background and the low
S/N prevent an unambiguous identification of the expected
absorption lines and only allows to put upper values regarding the
 presence or not of emission lines.

\item[(iv)] The analysis of XMM-Newton and Chandra observations shows 
that the X-ray
emission is elongated along the same direction as that defined by DS1, the
cluster center and DS2. 
The  dynamical groups G1, G2, G3, G5 and two rich 
galaxy clusters at $\simeq$ 40 Mpc (Abell~4041, Abell~S1151) from \a26\
also define a very similar direction.

\item[(v)] The X-ray temperature map shows the presence of a cool
core, i.e. a broad cool zone stretching roughly from north to south
and possibly suggests hotter regions towards the north east, south west
and north west, though error bars in these zones are large.  The
gas heating in these hotter regions could be due to shock fronts
produced along these directions by infalling material.  The general
elongation observed agrees with the proposed scenario in which diffuse sources
are probably concentrations of tidal debris and harassed matter
expelled from infalling galaxies by tidal stripping, and is therefore
expected to be found along the main infall directions. However,
this scenario is still speculative at this stage and would require
numerical simulations that are beyond the scope of this paper.

\end{description}

\begin{acknowledgements}

The authors thank Tatiana Ferraz Lagan\'a for deriving the X-ray
temperature profile, Francesco La Barbera for help on stellar
photometry, Nicola Napolitano for stimulating discussions, and the
referee for constructive remarks which helped improving the
presentation of the results. The authors also acknowledge useful
discussions with Genevi\`eve Soucail.  G.C. acknowledges financial
support from the EURO-3D Research Training Network (funded by the
European Commission under the contract No. HPRN-CT-2002-00305),
C.A. and F.D. acknowledge support from the PNC, F.D. and
G.B.L.N. acknowledge financial support from the CAPES/COFECUB, and JPK
from CNRS and Caltech.
This research has made use of the NASA/IPAC Extragalactic Database
which is operated by the Jet Propulsion Laboratory, 
California Institute of Technology, under contract 
with the National Aeronautics and Space Administration.

\end{acknowledgements}

\end{document}